\documentclass[fleqn,10pt]{wlscirep}
\usepackage[utf8]{inputenc}
\usepackage[T1]{fontenc}
\usepackage{import}
\usepackage{makeidx}
\usepackage{ulem}
\usepackage{symbols}
\usepackage{bm}
\usepackage{siunitx}
\usepackage{float}
\usepackage{stmaryrd}
\usepackage{textcomp}
\usepackage{collcell}
\usepackage{tabularx}
\usepackage{xcolor,colortbl}
\usepackage{booktabs}
\usepackage{lineno}
\usepackage{multirow}
\usepackage{algorithm}
\usepackage{algpseudocode}
\newcommand{\rev}[1]{{{\textcolor{black}{#1}}}}
\algnewcommand\algorithmicforeach{\textbf{for each}}
\algdef{S}[FOR]{ForEach}[1]{\algorithmicforeach\ #1\ \algorithmicdo}

\usepackage{lineno}

\title{Reduced order modeling for flow and transport problems with Barlow Twins self-supervised learning}

\author[1,2]{Teeratorn Kadeethum}
\author[3]{Francesco Ballarin}
\author[4]{Daniel O'Malley}
\author[5]{Youngsoo Choi}
\author[2,*]{Nikolaos Bouklas}
\author[1,*]{Hongkyu Yoon}

\affil[1]{Sandia National Laboratories, New Mexico, USA}
\affil[2]{Cornell University, New York, USA}
\affil[3]{Catholic University of the Sacred Heart, Brescia, Italy}
\affil[4]{Los Alamos National Laboratory, New Mexico, USA}
\affil[5]{Lawrence Livermore National Laboratory, California, USA}

\affil[*]{nbouklas@cornell.edu, hyoon@sandia.gov}

\begin{abstract}

We propose a unified data-driven reduced order model (ROM) that bridges the performance gap between linear and nonlinear manifold approaches. Deep learning ROM (DL-ROM) using deep-convolutional autoencoders (DC-AE) has been shown to capture nonlinear solution manifolds but fails to perform adequately when linear subspace approaches such as proper orthogonal decomposition (POD) would be optimal. Besides, most DL-ROM models rely on convolutional layers, which might limit its application to only a structured mesh. The proposed framework in this study relies on the combination of an autoencoder (AE) and Barlow Twins (BT) self-supervised learning, where BT maximizes the information content of the embedding with the latent space through a joint embedding architecture. Through a series of benchmark problems of natural convection in porous media, BT-AE performs better than the previous DL-ROM framework by providing comparable results to POD-based approaches for problems where the solution lies within a linear subspace as well as DL-ROM autoencoder-based techniques where the solution lies on a nonlinear manifold; consequently, bridges the gap between linear and nonlinear reduced manifolds. We illustrate that a proficient construction of the latent space is key to achieving these results, enabling us to map these latent spaces using regression models. The proposed framework achieves a relative error of 2\% on average and 12\% in the worst-case scenario (i.e., the training data is small, but the parameter space is large.). We also show that our framework provides a speed-up of $7 \times 10^{6}$ times, in the best case, and $7 \times 10^{3}$ times on average compared to a finite element solver. Furthermore, this BT-AE framework can operate on unstructured meshes, which provides flexibility in its application to standard numerical solvers, on-site measurements, experimental data, or a combination of these sources.

\end{abstract}
\begin{document}

\flushbottom
\maketitle
%
%
\thispagestyle{empty}


\section*{Introduction}

A reduced order model (ROM) is devised to provide an acceptable accuracy while utilizing a much lower computational cost compared to the full order model (FOM) \cite{hesthaven2016certified}. In recent years, a non-intrusive or data-driven ROM approach has grasped attention because (1) it has a straightforward implementation (i.e., does not require any modifications of FOM), (2) it easily lends itself to different kinds of physical problems, and (3) it allows for more stable and much faster prediction than intrusive ROM for nonlinear problems \cite{xiao2015non1,xiao2015non2,hesthaven2018non,fresca2021comprehensive,kadeethum2021nonTH,ahmed2021nonlinear}. Traditionally, proper orthogonal decomposition (POD) is used as a data compression tool (i.e., linear subspace approach), which is the optimal way to construct the linear reduced manifolds. However, POD-based solutions on a linear subspace are often restrictive for highly nonlinear problems where reduced spaces lie in nonlinear manifolds. More recently, nonlinear compression using autoencoder-based deep learning (DL) architectures or nonlinear manifold approach \cite{fresca2021comprehensive,kadeethum2021nonTH,kim2021fast,kim2020efficient} has been suggested to reconstruct these nonlinear manifolds, resulting in generic and more refined predictive capabilities than linear subspace approaches for nonlinear problems. Recent extensive comparisons, however, show a performance deficit for DL-ROM approaches in some cases \cite{kadeethum2021nonTH}. \par 

Kadeethum et al. \cite{kadeethum2021nonTH} illustrate that there are two essential issues for DL-ROM. First, the nonlinear approach outperforms its linear counterpart in specific settings (e.g., boundary conditions and domain geometry), but the opposite can occur in other settings. This is because POD provides the optimal data compression in a linear subspace for the problems with fast-decaying Kolmogorov's n-width that measures the degree of accuracy by n-dimensional linear subspaces \cite{chatterjee2000introduction,willcox2002balanced,choi2020sns,kim2021efficient}. Therefore, the DL-ROM approach could not exceed the level of POD accuracy for problems that naturally lie within linear manifolds. However, for problems with slowly decaying Kolmogorov's width, the nonlinear manifold approach outperforms the linear subspace one. Even though the authors hypothesize that a visual comparison between principal component analysis (PCA) and t-Distributed Stochastic Neighbor Embedding (t-SNE) could indicate which method will perform better before employing any specific compression strategy, there is no unified model that could be used across problem settings without an extensive case-based hyperparameter search. Second, although the nonlinear approach excels in very complex settings, it relies on convolutional operators, hindering its application for unstructured meshes and limiting DL-ROM approaches to less practical problems. Hence, these limitations in DL-ROM methods need to be resolved and tested with varying degrees of complex problems. \par

Convection in porous media is an important process in various applications in natural and engineered environments (e.g., biomedical engineering, multiphase flow in the subsurface, seawater intrusion, geothermal energy, and storage of nuclear and radioactive waste) \cite{taron2009thermal,nick2013reactive,zheng2002applied,rutqvist2005numerical}. As the media temperature and composition (fluid concentration) are altered, the dynamics of fluid density and viscosity variations could drive the flow field through flow instabilities \cite{nield2006convection,park2021microfluidic}. The gravity-driven flow problem is usually characterized by Rayleigh number ($\mathrm{Ra}$) in which if the $\mathrm{Ra}$ is low, the flow field is laminar, while if the $\mathrm{Ra}$ is high, the flow turns into a turbulent regime. In cases where the driving force is strong enough (very high $\mathrm{Ra}$), the flow might also exhibit fingering behavior \cite{davison2012pore}.


Numerical simulation of gravity-driven flow in porous media has been a subject of extensive research. Notable examples of full order model (FOM) include: (1) TOUGH software suite, which includes multi-dimensional numerical models for simulating the coupled thermo-hydro-mechanical-chemical (THMC) processes in porous and fractured media \cite{pruess1987tough,rutqvist2017overview},
(2) SIERRA Mechanics, which has simulation capabilities for coupling thermal, fluid, aerodynamics, solid mechanics and structural dynamics \cite{bean2012sierra},
(3) PyLith, a finite-element code for modeling dynamic and quasi-static simulations of coupled multiphysics processes \cite{aagaard2008pylith},
(4) OpenGeoSys project, which is developed mainly based on the finite element method using object-oriented programming THMC processes in porous media \cite{kolditz2012opengeosys},
(5) IC-FERST, a reservoir simulator based on control-volume finite element methods and dynamic unstructured mesh optimization \cite{obeysekara2018modelling},
(6) DYNAFLOW\texttrademark, a nonlinear transient finite element analysis platform \cite{prevost1983dynaflow},
(7) DARSim, multiscale multiphysics finite volume based simulator \cite{hosseinimehr2020adaptive},
(8) the CSMP, an object-oriented application program interface, for the simulation of complex geological processes, e.g.\ THMC, and their interactions  \cite{matthai2007numerical},
and
(9) PorePy, an open-source modeling platform for multiphysics processes in fractured porous media \cite{keilegavlen2019porepy}. \rev{In this study, we utilize the FOM developed in the previous works, a locally conservative mixed finite element framework for coupled hydro-mechanical–chemical processes in heterogeneous porous media \cite{kadeethum2020finite,kadeethum2021locally} in which interior penalty enriched Galerkin and mixed finite element are employed.} This FOM, however, is computationally expensive for two reasons. The first one is the problem of interest is highly nonlinear; hence, it takes more nonlinear iterations to converge. The second reason is to satisfy the Courant–Friedrichs–Lewy (CFL) condition, the FOM needs to march through many intermediate time-steps to reach the time-steps of interest \cite{diersch1988finite,frolkovivc2000numerical,kolditz1998coupled}.

Kadeethum et al. \cite{kadeethum2021nonTH} propose a data-driven reduced order model (ROM) that can reduce computation cost while maintaining an acceptable accuracy for natural convection in porous media problems. The model is applicable to parameterized problems \cite{carlberg2018conservative,ballarin2019pod,venturi2019weighted,hesthaven2016certified,choi2019space,copeland2022reduced,hoang2021domain,kim2021efficient}, depending on a set of parameters ($\bm{\mu}$) which could correspond to physical properties, geometric characteristics, or boundary conditions. This model sequentially follows (1) the offline and (2) online stages \cite{hesthaven2016certified,choi2021space}. The offline stage begins with initializing a set of input parameters,  which we call a training set. Then the FOM is solved corresponding to each member in the training set (in the following, we will refer to the corresponding solutions as snapshots). Either linear, relying on POD, \cite{hesthaven2018non,kadeethum2021non} or nonlinear compression, depending on deep convolutional autoencoder (DL-AE or DL-ROM) \cite{fresca2021comprehensive,kadeethum2021nonTH,kim2021fast}, is then used to compress FOM snapshots to produce basis functions that span reduced spaces of very low dimensionality, but still guarantee accurate reproduction of the snapshots \cite{decaria2020artificial,cleary1984data}. The ROM can then be solved during the online stage for any new value of $\bm{\mu}$ by seeking an approximated solution in the reduced space. \par

In this work, we propose a unified data-driven ROM using a combination of Barlow Twins (BT) self-supervised learning and an autoencoder (BT-AE) that bridges the performance gap between linear and nonlinear manifold approaches. In particular, we use BT self-supervised learning to maximize the information content of the embedding with the latent space through a joint embedding architecture \cite{zbontar2021barlow}. 
With four different example cases that span the degree of complexity to cover both linear and nonlinear problems, a comparison of the proposed BT-AE framework with both linear (POD) and nonlinear (DL-AE) ROM approaches is conducted to demonstrate the performance of the unified data-driven ROM framework that (1) excels in all test cases (whether the solution can be captured in a linear or nonlinear manifold) and (2) operates on either structured or unstructured meshes. Importantly, this model is fully data-driven; it could be trained by data produced by FOM, on-site measurement, experimental data, or a combination of them. This characteristic can provide flexibility across the spectrum in more complex problems. Since it is not limited by the Courant–Friedrichs–Lewy condition for conventional FOMs, it could deliver quantities of interest at any given time contrary to the FOM \cite{kadeethum2021nonTH}. \par


\section*{Results}

\subsection*{Data generation}

\begin{figure}[!ht]
  \centering
    \includegraphics[width=16.5cm,keepaspectratio]{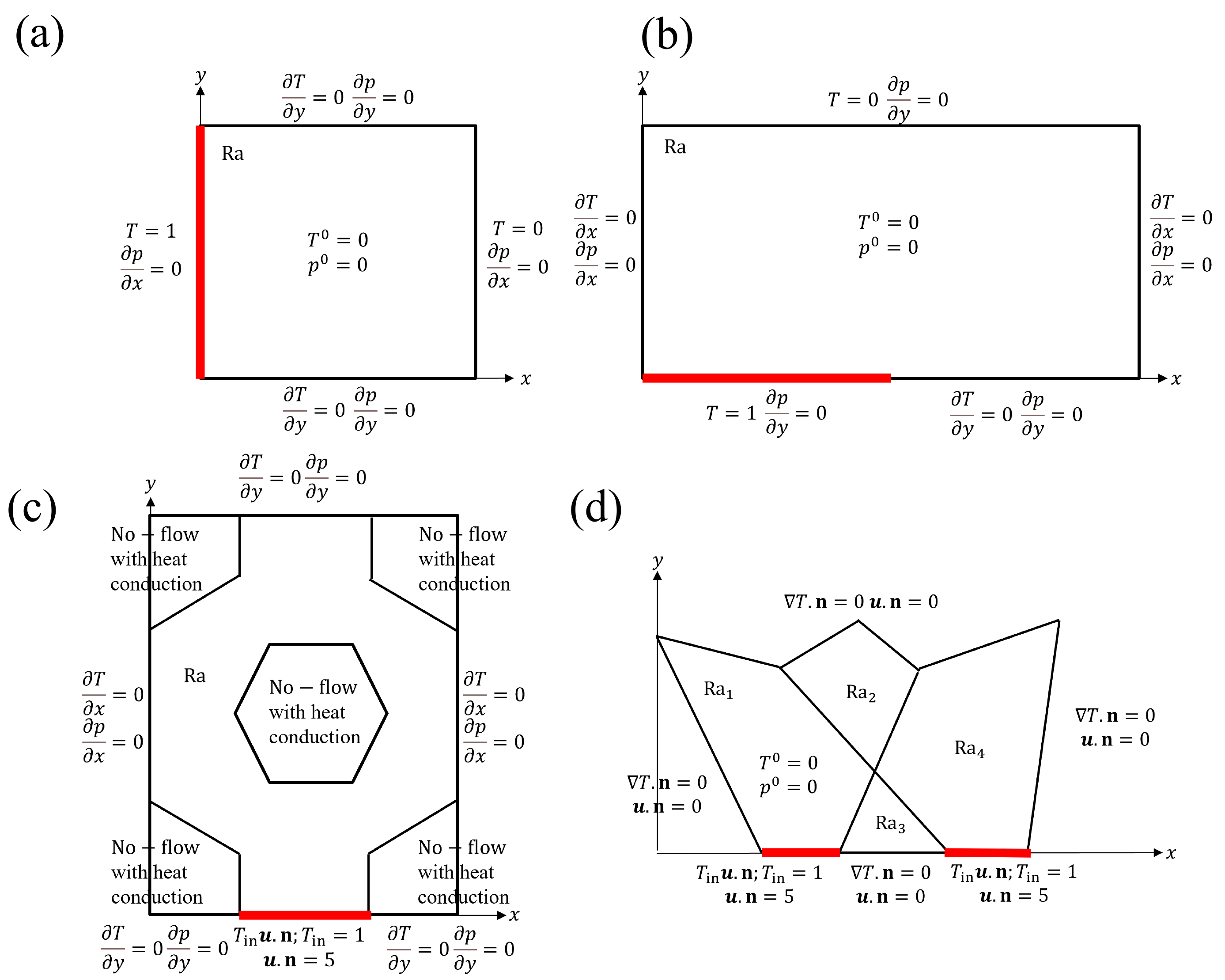}
  \caption{Domain and boundary conditions for (a) Example 1 (heating from the left boundary), (b) Example 2 (Elder problem), (c) Example 3 (unit cell of micromodel), and (d) Example 4 (modified Hydrocoin with four subdomains). The red line indicates the region of the boundary where the temperature is elevated.}
  \label{fig:geo}
\end{figure}

We present a summary of all geometries and boundary conditions we use in Figure \ref{fig:geo}. In short, Examples 1, 2, and 3 represent cases where $\bm{\mu}$ is a scalar quantity, namely $\mathrm{Ra}$, while Example 4 illustrates a case where $\bm{\mu}$ is a four-dimensional vector, composed of $\mathrm{Ra_1}$, $\mathrm{Ra_2}$, $\mathrm{Ra_3}$, and $\mathrm{Ra_4}$. The information of each example is presented in Table \ref{tab:main_info}. We note that $\mathrm{M}_\mathrm{validation}$ and $\mathrm{M}_\mathrm{test}$ represent the number of the validation and testing sets with varying Rayleigh number ($\mathrm{Ra}$), respectively (Table \ref{tab:main_info}). Due to time dependence, the total number of training, validation, and test samples is the product of $\mathrm{M}$ and $N^t$ with varying $N^t$ ranges. Specifically the validation samples, $\mathrm{M}_\mathrm{validation} N^t$, is determined by $\mathrm{M}_\mathrm{validation} N^t = 0.1\mathrm{M}N^t$ by randomly sampling 10\% of the sum of training/validation sets ($\mathrm{M}N^t$).  \par



\begin{table}[!ht]
\centering
\caption{Summary of main information for each example.}
\begin{tabular}{|l|c|c|c|c|l|}
\hline
                            & \textbf{Example 1}                      & \textbf{Example 2}                        & \textbf{Example 3} & \textbf{Example 4}           & \multicolumn{1}{c|}{\textbf{remark}}               \\ \hline
$\mathrm{M}$                & 40                                      & 40     & 40                                    & 81                          & training set for the parameter space $\bm{\mu}$                                      \\ \hline
$\mathrm{M}_\mathrm{validation} N^t$              & 10\% of $\mathrm{M} N^t$               & 10\% of $\mathrm{M} N^t$                         & 10\% of $\mathrm{M} N^t$                                        & 10\% of $\mathrm{M} N^t$                           & validation set - randomly select from $\mathrm{M} N^t$                                      \\ \hline
$\mathrm{M}_\mathrm{test}$              & 10            & 10                           & 10                                        & 10                           & test set for the parameter space $\bm{\mu}_{\mathrm{test}}$                                          \\ \hline

$\mathrm{M} N^t$            & 16802                                    & 36110               & 44354                        & 90175                        & total training/validation data                 \\ \hline

$\mathrm{M}_\mathrm{test} N^t$            & 3260                                    & 8951               & 11238                        & 11432                        & total testing data                 \\ \hline

$N^t$ range                 & {[}226, 477{]}                          & {[}790, 1010{]}     & {[}951, 1265{]}                        & {[}907, 1280{]}               & for training, validation, and test sets            \\ \hline
$N_{h}^T$                   & 7110                                    & 9600            & 17064                          & 11382                         & degree of freedom (DOF): $T_h$                                        \\ \hline
$t$ range                 & {[}0.0, 0.1{]}                          & {[}0.0, 0.1{]}                           & {[}0.0, 0.1{]}     & {[}0.0, 0.1{]}            & $\left[t^{0}, t^{N}\right]$           \\ \hline

\multirow{4}{*}{$\bm{\mu}$} & \multirow{4}{*}{$\mathrm{Ra} \in [40, 80]$} & \multirow{4}{*}{$\mathrm{Ra} \in [350, 450]$} & \multirow{4}{*}{$\mathrm{Ra} \in [350, 450]$} & $\mathrm{Ra_1} \in [350, 450]$ & \multirow{4}{*}{only Example 4 has four parameters} \\ \cline{5-5}
                   &                    &                    &                    & $\mathrm{Ra_2} \in [350, 450]$ &                    \\ \cline{5-5}
                   &                    &                    &                    & $\mathrm{Ra_3} \in [350, 450]$ &                    \\ \cline{5-5}
                   &                    &                    &                    & $\mathrm{Ra_4} \in [350, 450]$ &                    \\ \hline
\end{tabular}
\begin{flushleft} It is noted that the final total training samples are $0.9\mathrm{M} N^t$ because we allocate 10\% of the training samples for the validation set. The total of testing data is $\mathrm{M_{test}} N^t$. We want to emphasize that our $N^t$ is not constant, but it is a function of $\bm{\mu}$. To elaborate, the higher $\mathrm{Ra}$ value will result in the higher $N^t$ to satisfy CFL condition.  \end{flushleft}

\label{tab:main_info}
\end{table}

The summary of each model, including the subspace dimension and compression method, is presented in Table \ref{tab:naming}. The detailed description of POD, AE, and DC-AE models is provided in Kadeethum et al. \cite{kadeethum2021nonTH}, and our newly developed BT-AE models are described in the \textbf{Methodology} section. In short, for POD models, we use proper orthogonal decomposition as a compression tool. The AE models use an autoencoder as a compression method. We employ a deep convolutional autoencoder to compress our training snapshots ($\mathrm{M} N^t$) for DC-AE models. The BT-AE models utilize a combination of an autoencoder and Barlow Twins self-supervised learning in their compression procedure. For the POD models, linear compression, subspace dimension refers to the number of reduced basis or $\mathrm{N}$ as well as the number of intermediate reduced basis or $\mathrm{N_{int}}$. We assume $\mathrm{N} = \mathrm{N_{int}}$ for all models for simplicity. The subspace dimension is the number of latent space ($\mathrm{Q}$) for the nonlinear compression, AE, DC-AE, and BT-AE models. \par

\begin{table}[!ht]
\centering
\caption{Summary of naming for each model.}
\begin{tabular}{|c|c|c|c|}
\hline
\textbf{model name} & \textbf{compression} & \textbf{subspace dimension} & \textbf{compression techniques} \\ \hline
POD 16 RB           & linear               & 16                          & proper orthogonal decomposition                   \\ \hline
POD 50 RB           & linear               & 50                          & proper orthogonal decomposition                   \\ \hline
POD 100 RB          & linear               & 100                         & proper orthogonal decomposition                   \\ \hline
POD 500 RB          & linear               & 500                         & proper orthogonal decomposition                   \\ \hline
AE 4 Q       & nonlinear            & 4                           & autoencoder            \\ \hline
DC-AE 4 Q       & nonlinear            & 4                           & deep convolutional autoencoder             \\ \hline
BT-AE 4 Q         & nonlinear            & 4                           &  Barlow twins + autoencoder                 \\ \hline
AE 16 Q      & nonlinear            & 16                          & autoencoder            \\ \hline
DC-AE 16 Q      & nonlinear            & 16                          & deep convolutional autoencoder             \\ \hline
BT-AE 16 Q        & nonlinear            & 16                          & Barlow twins + autoencoder                   \\ \hline
AE 256 Q     & nonlinear            & 256                         & autoencoder            \\ \hline
DC-AE 256 Q     & nonlinear            & 256                         & deep convolutional autoencoder            \\ \hline
BT-AE 256 Q       & nonlinear            & 256                         & Barlow twins + autoencoder                  \\ \hline
\end{tabular}
\begin{flushleft} Details of POD, AE, DC-AE models are provided in Kadeethum et al. \cite{kadeethum2021nonTH}.  \end{flushleft}
\label{tab:naming}
\end{table}

\subsection*{Comparisons of BT-AE with POD, AE, and DC-AE models in simple domains}

We first compare the BT-AE model accuracy (for different numbers of $\mathrm{Q}$) with the models developed by Kadeethum et al. \cite{kadeethum2021non, kadeethum2021nonTH} (i.e., POD, AE, and DC-AE models) in relatively simple model domains. Example 1 illustrates a case where a linear manifold is optimal, while Example 2 presents a case where a nonlinear manifold is optimal. The results of POD, AE, and DC-AE models presented in Kadeethum et al. \cite{kadeethum2021nonTH} demonstrated that the POD-based and DL-ROM approaches are more suitable for the linear and nonlinear manifold problems, respectively, and they are used in this manuscript to evaluate the performance of BT-AE models.

\subsubsection*{Example 1: Heating from the left boundary}

The geometry and boundary conditions are shown in Figure \ref{fig:geo}a, and we adopt this example from Zhang et al. and Kadeethum et al. \cite{zhang2016mixed,kadeethum2021nonTH}. This example represents a case where its fluid flow is driven by buoyancy as the fluid is heated on the left side of the domain. The fluid then flows upwards and rotates to the right side of the domain. We set $\bm{\mu} = (\mathrm{Ra})$, and its admissible range of variation is $[40.0, 80.0]$, see Table \ref{tab:main_info}. For the training set, we use $\mathrm{M} = 40$, which lead to, in total, $\mathrm{M} N^t = 16802$ training data points. \par

We present the test case results of the BT-AE model (BT-AE 16 Q) as supplimental information (SI-Animation-Example 1). The difference between solutions produced by the FOM and ROM (DIFF) is calculated by
\begin{equation}\label{eq:diff}
\operatorname{DIFF}_\varphi(t^k, \bm{\mu}_\mathrm{test}^{(i)})= \left|\varphi_h(\cdot; t^k, \bm{\mu}_\mathrm{test}^{(i)}) - \widehat{\varphi}_h(\cdot; t^k, \bm{\mu}_\mathrm{test}^{(i)})\right|
\end{equation}

\noindent
where $\varphi_h$ is a finite-dimensional approximation of the set of primary variables corresponding to velocity, pressure, and temperature fields. $\widehat{\varphi}_h$ is an approximation of $\varphi_h$ produced by the ROM. Thus, $\varphi_h(\cdot; t^k, \bm{\mu}_\mathrm{test}^{(i)})$ and $\widehat{\varphi}_h(\cdot; t^k, \bm{\mu}_\mathrm{test}^{(i)})$ represent $\varphi_h$ and $\widehat{\varphi}_h$ at all space coordinates (i.e., evaluations at each DOF) at time $t^k$ with input parameter $\bm{\mu}_\mathrm{test}^{(i)}$, respectively. Note that we only present the results of the temperature field. Hence, $\varphi_h$ and $\widehat{\varphi}_h$ represent $T_h$ and $\widehat{T}_h$, respectively. From SI-Animation-Example 1, we observe that BT-AE 16 Q provides a reasonable approximation of the temperature field. 

\par

\begin{figure}[!ht]
  \centering
    \includegraphics[width=11.0cm,keepaspectratio]{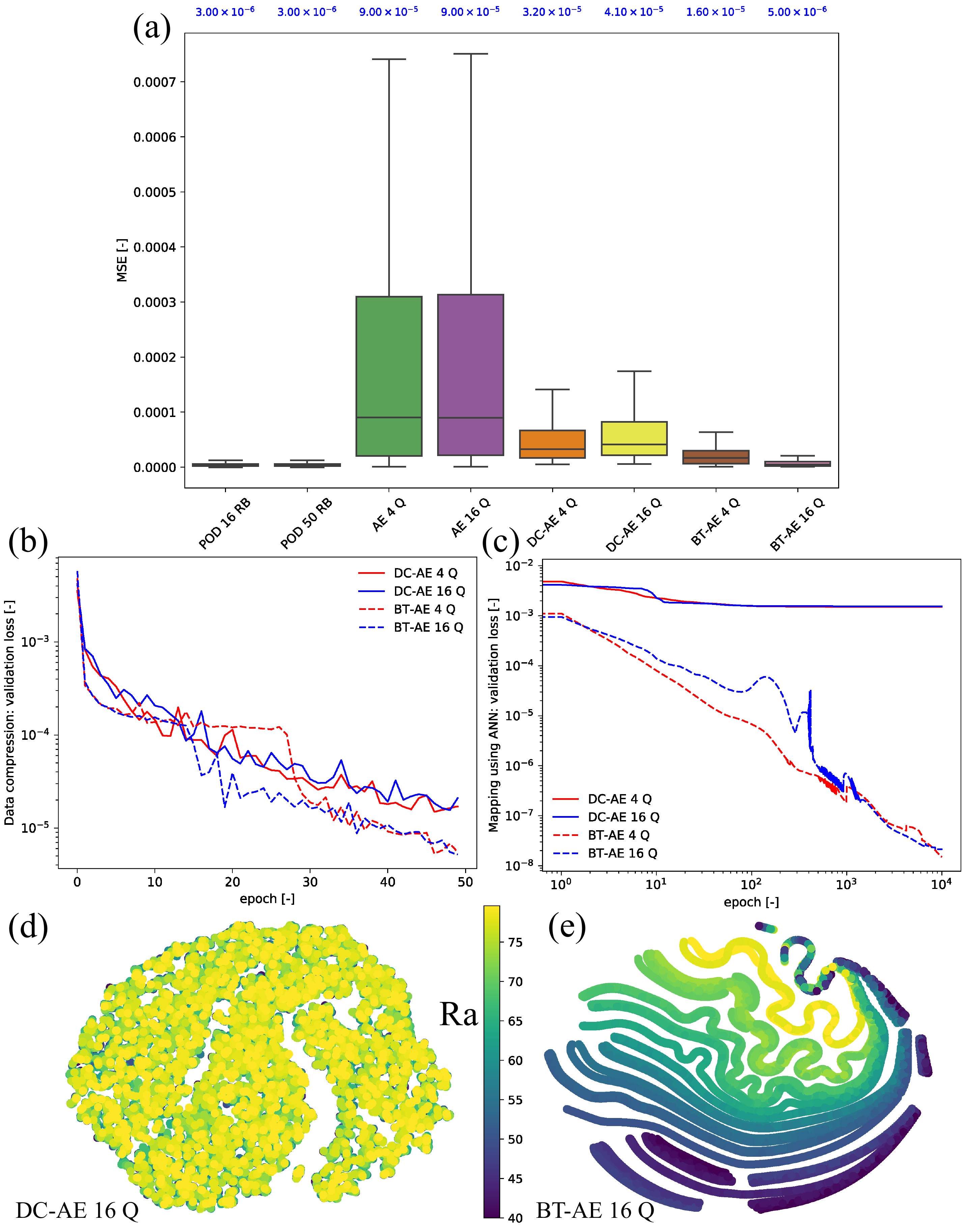}
  \caption{Example 1 - results: (a) mean squared error (MSE) of each model (please refer to Table \ref{tab:naming}), and blue texts represent a mean value of the box plots - here we show that BT-AE 16 gives performance similar to POD-based approaches, but AE and DC-AE models don't, (b) data compression loss for validation set (Equation \eqref{eq:loss_ae}), (c) mapping using ANN loss for validation set (Equation \eqref{eq:loss_ann_nonlinear}), (d) latent space plot of DC-AE 16 Q model, and (e) latent space plot of BT-AE 16 Q model. Latent space plots are constructed using t-Distributed Stochastic Neighbor Embedding (t-SNE). Different colors represent each value of $\mathrm{Ra}$ value. We calculate the t-SNE plots using Scikit-Learn package using its default setting and perplexity of 15.}
  \label{fig:hfs_results}
\end{figure}

The results of Example 1 is presented in Figure \ref{fig:hfs_results}. In Figure \ref{fig:hfs_results}a, The performance of the different models (Table \ref{tab:naming}) is evaluated with the mean square error ($\mathrm{MSE}_\varphi(:, \bm{\mu}_\mathrm{test}^{(i)})$) of the test cases defined as follows
\begin{equation}\label{eq:validation_mse}
    {\mathrm{MSE}_\varphi(:, \bm{\mu}_\mathrm{test}^{(i)}) :=\frac{1}{N^t} { \sum_{k=0}^{N^t} \left| \varphi_h(\cdot; t^k, \bm{\mu}_\mathrm{test}^{(i)}) - \widehat{\varphi}_h(\cdot; t^k, \bm{\mu}_\mathrm{test}^{(i)})\right|_{\varphi}^2}}
\end{equation}

\noindent
where $\mathrm{MSE}_\varphi(:, \bm{\mu}_\mathrm{test}^{(i)})$ represents the MSE values of all $t$ for each $\bm{\mu}_\mathrm{test}^{(i)}$. The $\mathrm{MSE}$ results show that BT-AE models perform better than AE and DC-AE models. Besides, BT-AE 16 Q delivers similar $\mathrm{MSE}$ results to those of the POD models. In contrast to the findings presented in Kadeethum et al. \cite{kadeethum2021nonTH} where the linear compression (POD) outperforms nonlinear compression (AE and DC-AE), BT-AE models in this study could perform similar to the POD models. To be accurate, BT-AE models still underperform, but errors are comparable. \par

We then investigate how the performance of BT-AE models compares to DC-AE. First, we examine the data compression loss of the validation set (see Equation \eqref{eq:loss_ae}) which is presented in Figure \ref{fig:hfs_results}b. From this Figure, the data compression losses of BT-AE models are slightly better than those of the DC-AE models. Subsequently, we illustrate the mapping using ANN loss of the validation set, see Equation \eqref{eq:loss_ann_nonlinear}, in Figure \ref{fig:hfs_results}c. From Figure \ref{fig:hfs_results}c, we observe that the mapping losses of the BT-AE models are six orders of magnitude less than those of the DC-AE models. This behavior shows that the BT-AE's latent spaces are easier to be mapped (i.e., ANN loss of the validation set for the BT-AE mapping is much lower than that of the DC-AE.). This speculation is explained by Figures \ref{fig:hfs_results}d-e, using a t-Distributed Stochastic Neighbor Embedding (t-SNE) plot. From Figure \ref{fig:hfs_results}d, one could see that all latent variables of DC-AE 16 Q blend (i.e., you cannot differentiate among cases with different $\mathrm{Ra}$ values.). The latent variables of the BT-AE 16 Q model, on the other hand, shown in Figure \ref{fig:hfs_results}e, behave in a much better structure (i.e., we can differentiate among cases with different $\mathrm{Ra}$ values.). \par

\subsubsection*{Example 2: Elder problem}

The Elder problem \cite{elder1967transient} is a significantly more complicated and ill-posed problem \cite{elder1967transient, simpson2003theoretical}. High $\mathrm{Ra}$ numbers considered in this case may cause the flow instability to be fingering behavior. The domain and boundary conditions are presented in Figure \ref{fig:geo}b \cite{diersch2002variable, zhang2016mixed, kadeethum2021nonTH}. In short, the model domain is heated from the half of the bottom boundary (Figure \ref{fig:geo}b), and the flow is driven upwards by buoyancy force. We set $\bm{\mu} = (\mathrm{Ra})$, and its admissible range as $[350.0, 400.0]$ (Table \ref{tab:main_info}). Compared to Example 1, this higher range of $\mathrm{Ra}$ values affects the minimum and maximum $N^t$ as its range increases to $[790, 1010]$. \par

\begin{figure}[!ht]
  \centering
    \includegraphics[width=11.0cm,keepaspectratio]{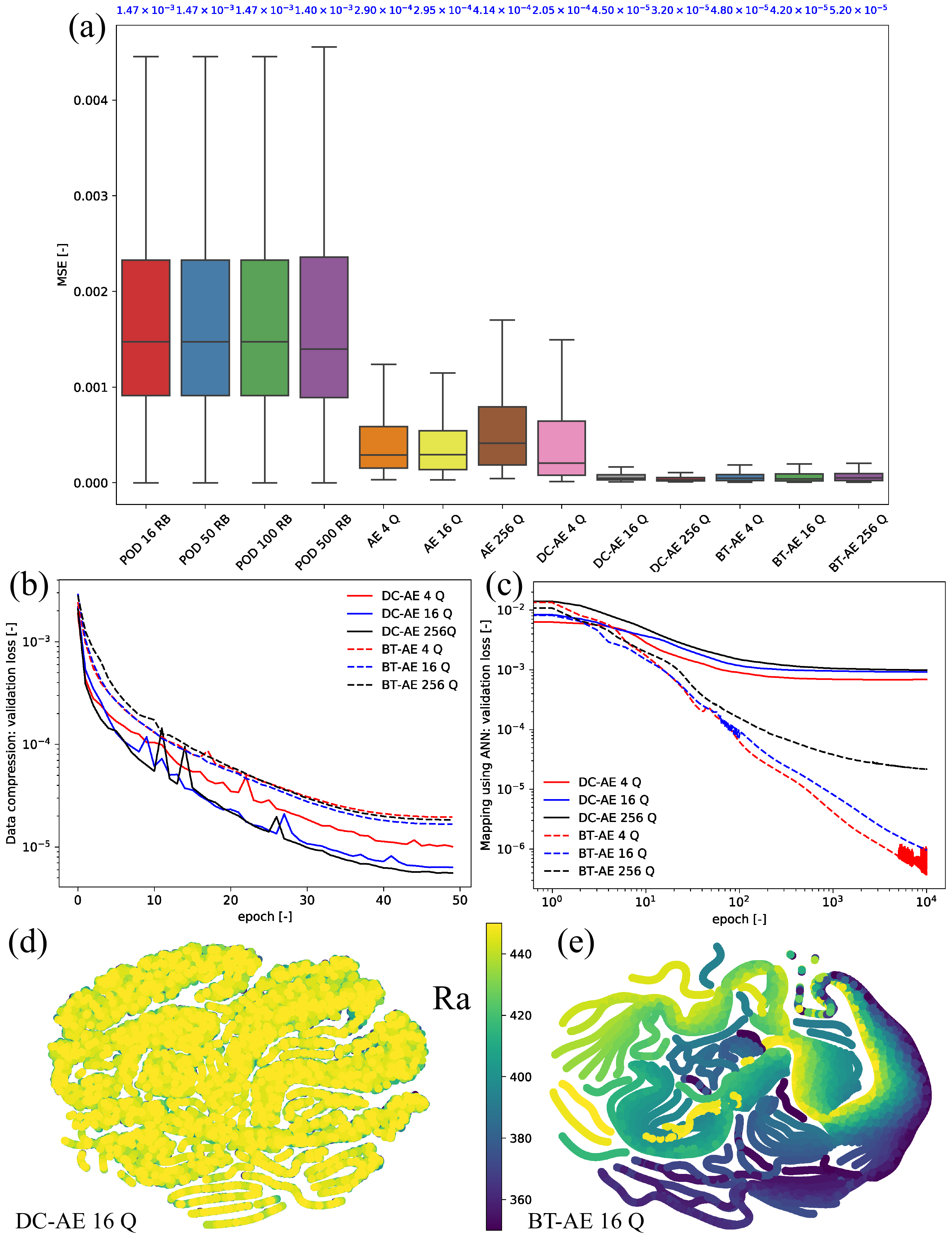}
  \caption{Example 2 - results: (a) mean squared error (MSE) of each model (please refer to Table \ref{tab:naming}), and blue texts represent a mean value of the box plots - here we show that BT-AE models provide performance similar to DC-AE models, but POD-based approaches and AE models don't, (b) data compression loss for validation set (Equation \eqref{eq:loss_ae}), (c) mapping using ANN loss for validation set (Equation \eqref{eq:loss_ann_nonlinear}), (d) latent space plot of DC-AE 16 Q model, and (e) latent space plot of BT-AE 16 Q model. Latent space plots are constructed using t-Distributed Stochastic Neighbor Embedding (t-SNE). Different colors represent each value of $\mathrm{Ra}$ value. We calculate the t-SNE plots using Scikit-Learn package using its default setting and perplexity of 15.}
  \label{fig:elder_results}
\end{figure}

The results of Example 2 are presented in Figure \ref{fig:elder_results}. From Figure \ref{fig:elder_results}a, we observe that all the models using nonlinear compression (AE, DC-AE, and BT-AE) perform better than the linear compression (POD). Furthermore, the BT-AE model accuracy is comparable to that of the DC-AE models. However, the BT-AE model results seem to be insensitive to the number of $\mathrm{Q}$, while the DC-AE model results are affected by the number of $\mathrm{Q}$ (i.e., the DC-AE 16 Q and DC-AE 256 Q are more accurate than the DC-AE 4 Q.). We also present the results of the test cases for the BT-AE 16 Q model in the supplemental animation (SI-Animation-Example 2). From these results, we observe that the BT-AE 16 Q model delivers a reasonable approximation of the solution $T_h$ (i.e., $\widehat{T}_h$).   \par 

We present the data compression loss of the validation set (Equation \eqref{eq:loss_ae}) in Figure \ref{fig:elder_results}b. In contrast to the ones shown in Figure \ref{fig:hfs_results}b, the DC-AE models have a slightly lower loss than that of the BT-AE models. We then investigate the ANN mapping loss (see Equation \eqref{eq:loss_ann_nonlinear}) of the validation set in Figure \ref{fig:elder_results}c. Similar to those presented in Figure \ref{fig:hfs_results}c, the BT-AE models have much lower mapping losses compared to those of the DC-AE models. Among the BT-AE models, BT-AE 256 Q has the highest value of ANN mapping loss, which is expected since it has the highest output dimension (i.e., we are mapping $t$ and $\bm{\mu}$ to $\bm{z}^{\mathrm{Q}}$). Again, we observe a much better structure of the BT-AE 16 Q latent space than the one from DC-AE 16 Q (see Figures \ref{fig:elder_results}d-e). To elaborate, the latent variables of the DC-AE 16 Q are overlapped to hinder us from differentiating among each case (different $\mathrm{Ra}$ values). The latent variables of the BT-AE, on the contrary, are structured in a way that one can clearly observe different parts that represent different $\mathrm{Ra}$ values as shown in Figures \ref{fig:elder_results}e. \par

\subsection*{Model performance of BT-AE models on complex geometries}

From Examples 1 and 2, we have observed that the BT-AE models could provide good results while operating on unstructured data. In this section, more challenging geometries which require an unstructured mesh for the FOM are evaluated with BT-AE models only since other methods are not suitable for unstructured mesh problems. \par

\subsubsection*{Example 3: Unit cell of micromodel}

Example 3 uses a unit cell of micromodel where a central part of honeycomb shape and four corners are impermeable for flow. Still, the heat can conduct through these five subdomains as presented in Figure \ref{fig:geo}c. Over the past decade, the micromodel has been used to study multiple coupled processes, including flow, reactive transport, bioreaction, and flow instability \cite{yoon2012pore, yoon2015lattice, yoon2019pore, yoon2014adaptation, davison2012pore, park2021microfluidic}. The flow is initiated from an influx at the bottom of the domain. This geometry is more complex than those utilized in Examples 1 and 2 (see Figures \ref{fig:geo}a-b). The higher temperature at the bottom surface (shown in red) alters a fluid density at the bottom, and subsequently, a buoyancy force drives the flow upwards from the bottom (shown in red) to the top of the domain. Five subdomains contain very low flow conductivity, but they can conduct heat. Again, we set $\bm{\mu} = (\mathrm{Ra})$ and its range as $[350.0, 400.0]$ (Table \ref{tab:main_info}). The range of $\mathrm{Ra}$ can also cause flow instability. We use $\mathrm{M} = 40$, $\mathrm{M}_\mathrm{validation}N^t = 10$\% of $\mathrm{M}N^{t}$, and $\mathrm{M}_\mathrm{test} = 10$. We have in total $\mathrm{M} N^t = 44354$ training data points. \par

\begin{figure}[!ht]
  \centering
    \includegraphics[width=11.0cm,keepaspectratio]{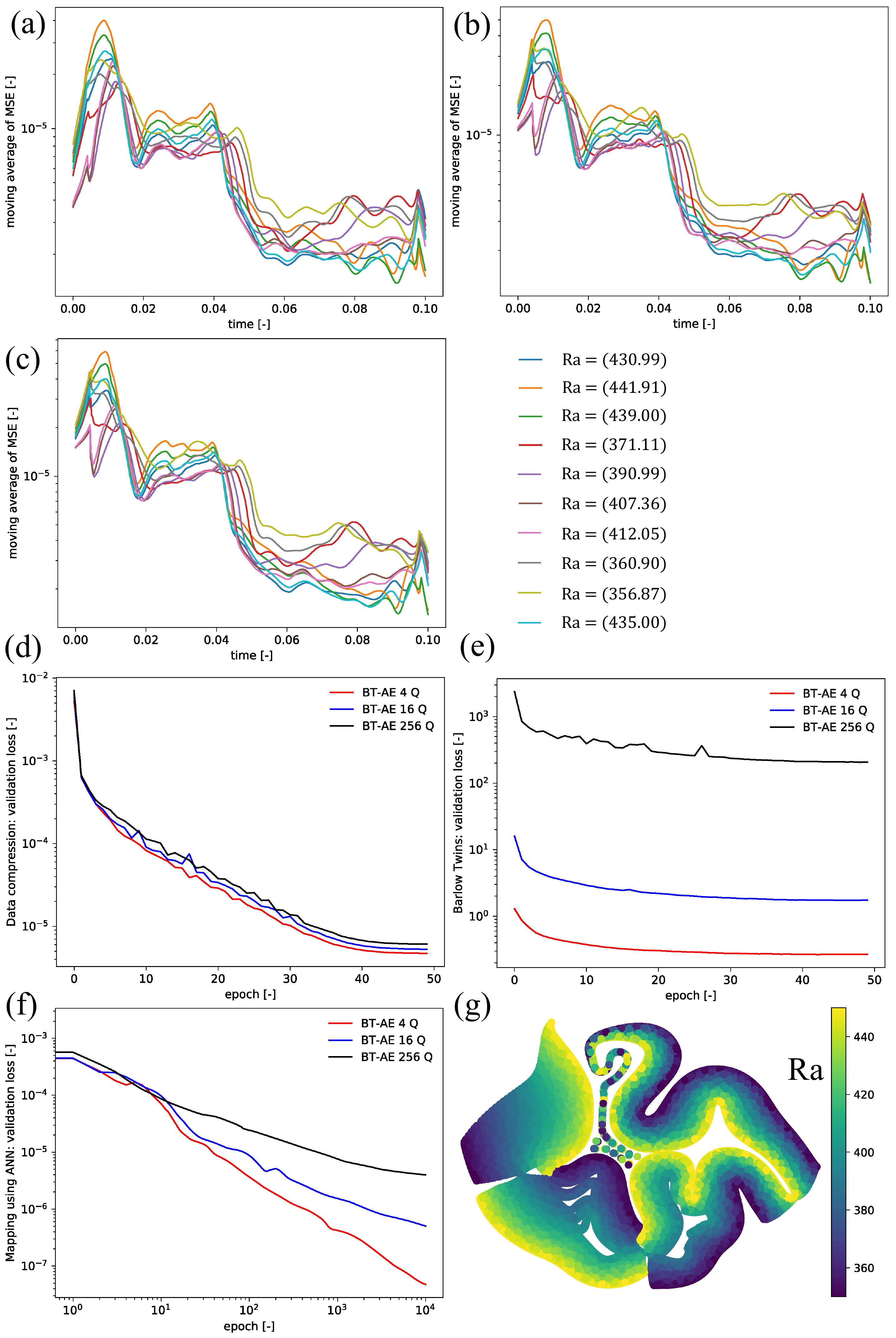}
  \caption{Example 3 results: the moving average (a window size of 50) of mean squared error (MSE) of (a) BT-AE 4 Q, (b) BT-AE 16 Q, (c) BT-AE 256 Q (please refer to Table \ref{tab:naming}). (d) Data compression loss for validation set (Equation \eqref{eq:loss_ae}), (e) Barlow Twins loss for validation set (Equation \eqref{eq:loss_bt}), (f) mapping using ANN loss for validation set (Equation \eqref{eq:loss_ann_nonlinear}), and (g) latent space plot of BT-AE 16 Q model. Latent space plots are constructed using t-Distributed Stochastic Neighbor Embedding (t-SNE). Different colors represent each value of $\mathrm{Ra}$ value. We calculate the t-SNE plots using Scikit-Learn package using its default setting and perplexity of 15.}
  \label{fig:micro_1mu}
\end{figure}

The summary of the Example 3 results is shown in Figure \ref{fig:micro_1mu}. For all test cases the MSE values over time in Figures \ref{fig:micro_1mu}a-c are in the range of $\approx 1 \times 10^{-5}$. The MSE values tend to decrease over time until the temperature field becomes a steady state. Besides, BT-AE models with different $\mathrm{Q}$ values provide approximately similar results (in line with our findings from Examples 1 and 2). The behavior infers that utilizing only a small number of latent spaces; the model can achieve the same level of accuracy as the one with a large number of latent spaces. This behavior is very beneficial because the mapping between parameter space and latent space becomes more manageable. We also present the results of the test cases for the BT-AE 16 Q model in the supplemental animation (SI-Animation-Example 3). Overall, BT-AE 16 Q delivers a reasonable approximation of the $T_h$ (i.e., DIFF results are low, and the relative error lies within 2 \%). \par

The data compression loss (Equation \eqref{eq:loss_ae}) is in the range of $\approx 1 \times 10^{-5}$ to $1 \times 10^{-6}$ (Figure \ref{fig:micro_1mu}d) which is similar to that of Example 1 (Figure \ref{fig:hfs_results}b), but slightly lower than that of the Example 2 (Figure \ref{fig:elder_results}b). The data compression loss seems to be invariant to $\mathrm{Q}$ values. We also present the Barlow Twins loss (Equation \eqref{eq:loss_bt}) in Figure \ref{fig:micro_1mu}e. We observe that the Barlow Twins loss increases with increasing the $\mathrm{Q}$ value as in Zbontar et al. \cite{zbontar2021barlow}. This can be explained that as the $\mathrm{Q}$ value grows larger, the cross-correlation matrix $\mathbf{C}^{T}\left(t, \bm{\mu}\right)$ becomes bigger, resulting in more members in Equations \eqref{eq:loss_I} and \eqref{eq:loss_RR}. As stated by Zbontar et al. \cite{zbontar2021barlow}, the absolute value of Equations \eqref{eq:loss_I} and \eqref{eq:loss_RR} is not as important as their trend. To elaborate this, in Figure \ref{fig:micro_1mu}e, all models (different $\mathrm{Q}$ values) reach their saturated points around 40 epochs, meaning that the minimization of Equations \eqref{eq:loss_I} and \eqref{eq:loss_RR} is completed.  \par

The mapping of the latent space using ANN loss (Equation \eqref{eq:loss_ann_nonlinear}) is presented in Figure \ref{fig:micro_1mu}f. Similar to Examples 1 and 2 (Figures \ref{fig:hfs_results}c and \ref{fig:elder_results}c), the mapping loss is in range of $\approx 1 \times 10^{-5}$ to $1 \times 10^{-7}$. The higher $\mathrm{Q}$ values, the mapping loss grows larger because there are more outputs to map. We present the latent space structure in Figure \ref{fig:micro_1mu}g (only for BT-AE 16 Q). Following the results shown in Figures \ref{fig:hfs_results}e and \ref{fig:elder_results}e, the latent structure of the BT-AE model has a good structure since we can differentiate among different $\mathrm{Ra}$ values. This behavior stems from the fact that the BT losses maximize the information content of the embedding with the latent space through a joint embedding architecture. \par


\subsubsection*{Example 4: Modified Hydrocoin with four subdomains}

Example 4 uses the hydrocoin problem \cite{inspectorate1987international, flemisch2018benchmarks} with the domain geometry shown in Figure \ref{fig:geo}d. In this example, the domain is subdivided into four subdomains with different $\mathrm{Ra}$ values (i.e., $\bm{\mu} = \left(\mathrm{Ra_1}, \mathrm{Ra_2}, \mathrm{Ra_3}, \mathrm{Ra_4}\right)$). The range of $\mathrm{Ra}$ values is $[350.0, 400.0]$. Similar to the previous examples, this $\mathrm{Ra}$ range causes fingering behavior as shown in the supplemental animation (SI-Animation-Example4). We use $\mathrm{M} = 81$, $\mathrm{M}_\mathrm{validation}N^t = 10$\% of $\mathrm{M}N^{t}$, and $\mathrm{M}_\mathrm{test} = 10$. We have in total $\mathrm{M} N^t = 90175$ training data points. We note that as we use $\mathrm{M} = 81 = 3^4$ equally spaced samples, for each parameter $\mathrm{Ra}_i$, $i = 1,2,3,4$, we only have three values. As an example, for $\mathrm{Ra_1}$ we only sample $\mathrm{Ra_1} = \left(350, 400, 450\right)$ for the training set. The same goes for $\mathrm{Ra_2}, \mathrm{Ra_3},$ and $\mathrm{Ra_4}$. As a result, training with relatively sparse samples of each parameter $\mathrm{Ra}_i$ makes it very challenging to obtain an accurate data-driven framework in general \cite{hesthaven2016certified,hesthaven2018non}. \par

\begin{figure}[!ht]
  \centering
    \includegraphics[width=11.0cm,keepaspectratio]{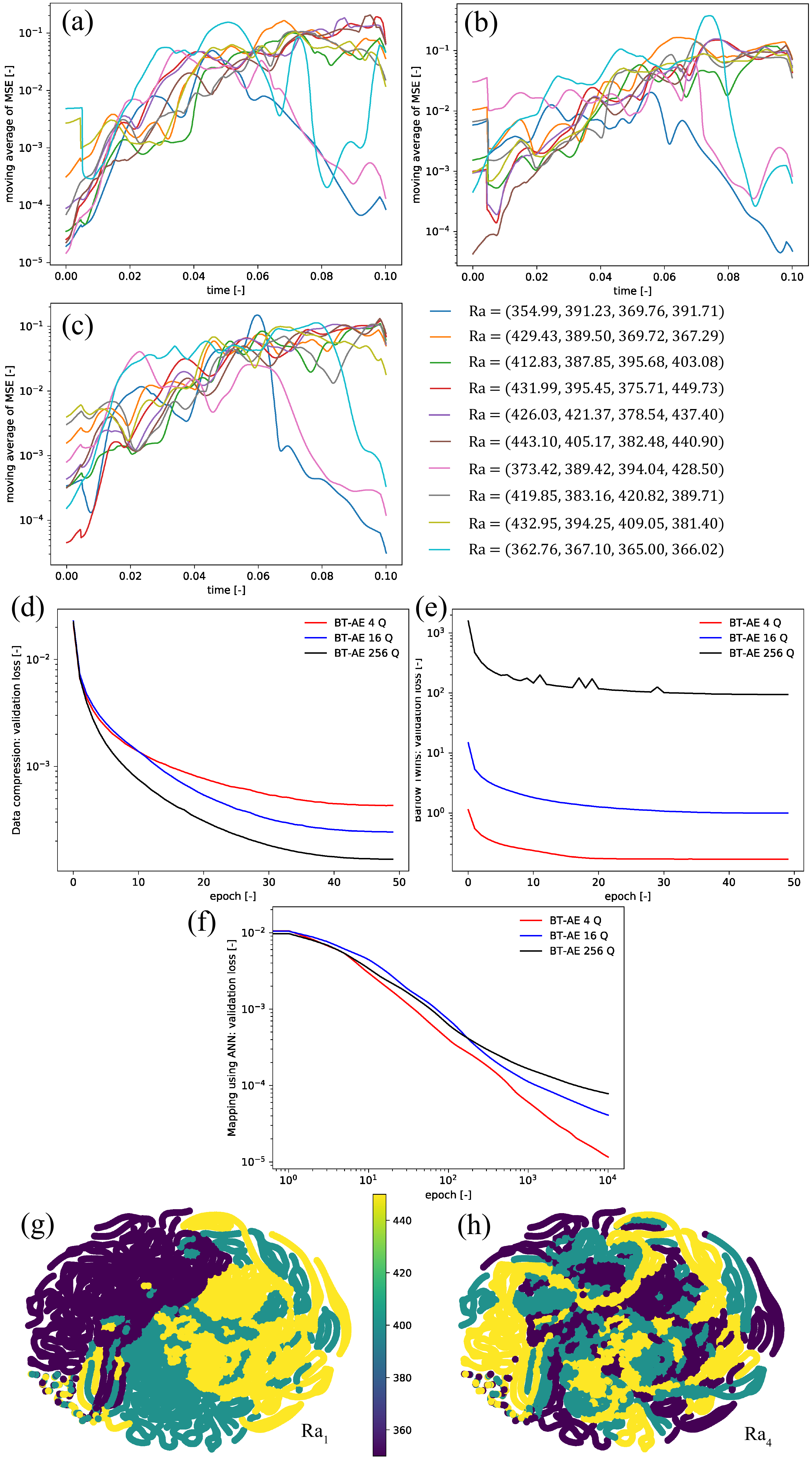}
  \caption{Example 4 results: the moving average (a window size of 50) of mean squared error (MSE) of (a) BT-AE 4 Q, (b) BT-AE 16 Q, (c) BT-AE 256 Q (please refer to Table \ref{tab:naming}). (d) Data compression loss for validation set (Equation \eqref{eq:loss_ae}), (e) Barlow Twins loss for validation set (Equation \eqref{eq:loss_bt}), (f) mapping using ANN loss for validation set (Equation \eqref{eq:loss_ann_nonlinear}), and latent space plot of BT-AE 16 Q model for (g) $Ra_1$ and (h) $Ra_4$. Latent space plots are constructed using t-Distributed Stochastic Neighbor Embedding (t-SNE). Different colors represent each value of $\mathrm{Ra}$ value. We calculate the t-SNE plots using Scikit-Learn package using its default setting and perplexity of 15.}
  \label{fig:hydro4mu_results}
\end{figure}

Even though this setting is very challenging, we still observe that the  BT-AE 16 Q delivers a reasonable approximation of the $T_h$ as seen in the supplemental animation (SI-Animation-Example4). The summary of the Example 4 results is shown in Figure \ref{fig:hydro4mu_results}. We present the MSE values as a function of time in Figure \ref{fig:hydro4mu_results}a-c. We can observe that the MSE values for all test cases are in the range of $\approx 1 \times 10^{-1}$ to $1 \times 10^{-5}$, which are significantly higher than those of Examples 1, 2, and 3. Moreover, the MSE values generally increase as we approach steady-state solutions, unlike the behaviors shown in Example 3. Again, BT-AE models with different $\mathrm{Q}$ provide approximately similar results (in line with our finding from Examples 1, 2, and 3). \par

The data compression loss (Equation \eqref{eq:loss_ae}) is in the range of $\approx 1 \times 10^{-2}$ to $1 \times 10^{-4}$ (Figure \ref{fig:hydro4mu_results}d), which is significantly higher than that of Examples 1, 2, and 3. This behavior illustrates that this example is the most challenging case for the BT-AE models. The data compression loss is the lowest for $\mathrm{Q} = 256$ and the highest for $\mathrm{Q} = 4$, but the difference is not critical. As shown in the Barlow Twins loss (Equation \eqref{eq:loss_bt}) in Figure \ref{fig:hydro4mu_results}e, the higher values of $\mathrm{Q}$ the larger Barlow Twins loss is (as we discussed in the previous example.). \par

The mapping of the latent space using ANN loss (Equation \eqref{eq:loss_ann_nonlinear}) is presented in Figure \ref{fig:hydro4mu_results}f. The mapping loss is in the range of $\approx 1 \times 10^{-4}$ to $1 \times 10^{-5}$, which is significantly higher than those of Examples 1, 2, and 3 (see Figures \ref{fig:hfs_results}c, \ref{fig:elder_results}c, and \ref{fig:micro_1mu}f). This behavior also contributes to the higher MSE values of the BT-AE models. We present the latent space structure (only for BT-AE 16 Q) in Figure \ref{fig:hydro4mu_results}g-h for $\mathrm{Ra_1}$ and $\mathrm{Ra_4}$, respectively. Since Example 4 has different $\mathrm{Ra}$ values in four subdomains, the differentiation of the latent space of individual $\mathrm{Ra}$ does not provide good solutions as each latent space of each subdomain might also be interconnected. \par

\section*{Discussion}

Recent developments in ML-based data-driven reduced order modeling (DL-ROM or DC-AE in this study) \cite{fresca2021comprehensive,kadeethum2021nonTH} have shown promising results in capturing parametrized solutions of systems of nonlinear equations. These models, however, rely on convolutional operators, which hinders the applicability of these models to complex geometries where an unstructured mesh is required for FOMs, as in Examples 3 and 4. Though we could utilize an autoencoder without convolutional layers, the model could not achieve the same level of accuracy as DL-ROM \cite{kadeethum2021nonTH}. Kadeethum et al. \cite{kadeethum2021nonTH} also illustrate that in a specific setting (simple geometry and boundary conditions), a linear compression approach using POD can outperform the DL-ROM model (Example 1). We have demonstrated that the autoencoder model through Barlow Twins self-supervised learning (BT-AE) could achieve the same accuracy as DL-ROM (Example 2 where POD models perform much worse than DL-ROM) by regularizing the latent space or nonlinear manifolds. Besides, it also yields optimal results in the case where the linear compression model outperforms the DL-ROM (Example 1). It means that the BT-AE model excels in all the test cases (Examples 1 and 2) while it still can operate on an unstructured mesh. This behavior has a significant advantage in scientific computing since most realistic problems require unstructured mesh representations. Besides, the BT-AE's performance is insensitive to the number of latent spaces, suggesting that with only a small number of latent spaces, the model can achieve the same level of accuracy as the one with a large number of latent spaces. This behavior is very beneficial because the mapping between parameter space and latent space becomes more manageable. \par

The computational time used to develop our ROM can be broken down into three primary parts: (1) generation of training data through FOM (the second step in Figure \ref{fig:model}), (2) training BT-AE (the third step in Figure \ref{fig:model}), and (3) mapping of $t$ and $\bm{\mu}$ to reduced subspace (the fourth step in Figure \ref{fig:model}). Each FOM model (corresponding to each set of $\bm{\mu}$ or $\mathrm{Ra}$ in this work) takes, on average, about two hours on AMD Ryzen Threadripper 3970X (4 threads). We note that our FOM utilizes the adaptive time-stepping; hence, each $\bm{\mu}^{(i)}$ may require a substantially different computational time. To elaborate, cases that have higher $\mathrm{Ra}$ usually have a smaller time-step ($N^t$ becomes larger), and subsequently, they require more time to complete.

The wall time used to train BT-AE is approximately 0.4 hours using a single Quadro RTX 6000. It is noted that this computational cost is much cheaper than that of the DC-AE model, taking around four to six hours \cite{kadeethum2021nonTH}. This is because DC-AE relies on convolutional layers, dropout, and batch normalization, which require much higher computational resources. The BT-AE, on the other hand, utilizes only a plain autoencoder. The BT-AE model is also cheaper than the POD model. However, we note that this may not be a fair comparison as we perform POD and BT-AE using different machines (i.e., our POD framework only works on CPU, but our BT-AE is trained using GPU). Please refer to Kadeethum et al. \cite{kadeethum2021nonTH} for detailed wall time comparisons among POD and DC-AE models. The mapping of $t$ and $\bm{\mu}$ to reduced subspace through artificial neural networks (ANN) takes around half an hour to one hour using a single Quadro RTX 6000. As mentioned in the Methodology section, we do not terminate the training of both BT-AE and mapping of $t$ and $\bm{\mu}$ to reduced subspace through ANN early, but rather use the model with the best validation loss through the final epochs. For example, we train for 50 epochs, but the model that offers the best validation loss might be the model at 20 epochs. However, the training time we report here is for 50 epochs. Thus, our training time provided here is considered conservative. \par

Even though the ROM training time is not trivial, it could provide a fast prediction during the online phase. Using AMD Ryzen Threadripper 3970X, the ROM takes approximately several milliseconds for a query of a pair of $t^{k}$ and $\bm{\mu}^{(i)}$. We also note that, as discussed previously, our ROM is needed to be trained on GPU for the problems at hand. Still, it could utilize CPU during an online time since we do not have to deal with back-propagation or optimization during the prediction time. On the contrary, one FOM simulation (for each $\bm{\mu}^{(i)}$ for all $t$ $\in$ $0=: t^{0}<t^{1}<\cdots<t^{N} := \tau$) takes about two hours. So, assuming that we query all $t$ similar to those of the FOM, ROM takes only a matter of several seconds. In practice, however, we might not need to evaluate all timestamps in $0=: t^{0}<t^{1}<\cdots<t^{N}:= \tau$ because the quantities of interest at the specific time may be more important. Since ROM is not bound by the CFL condition and can predict the quantities of interest at any specific time without intermediate computation, we could simply perform one query - $t^{N}$ and $\bm{\mu}^{(i)}$, resulting in saving computational time significantly. Our ROM could provide a speed-up of $7 \times 10^{6}$ at any specific time step for Example 2, and a speed-up of $7 \times 10^{3}$ to $7 \times 10^{6}$ for all examples considered in this work.



Our model is developed upon the data-driven paradigm, which is applicable to any FOM. Besides, it could be trained using data produced by FOM, on-site measurements, experimental data, or a combination among them. This characteristic provides flexibility, which intrusive approaches could not provide. The data-driven model, though, is usually hungry for training samples. We have illustrated that as the dimensionality of our parameter space grows, the model requires more training samples, or it will suffer by losing its accuracy significantly as in Example 4 compared to accurate prediction in  Example 3. We speculate that an adaptive sampling technique \cite{paul2015adaptive,vasile2013adaptive,choi2020gradient}, incorporating physical information \cite{raissi2019physics,kadeethum2020physics}, or including multimodal unsupervised training \cite{huang2018multimodal} might provide a resolution to this issue in the future work. \rev{Another gap in data-driven machine learning ROM is that a posteriori error is exceptionally challenging to quantify. An error estimator developed by Xiao \cite{xiao2019error} for linear manifolds could be adapted and extended to the nonlinear manifold paradigm. Additionally, epistemic uncertainty could also be quantified by adopting the ensemble technique proposed by Jacquier et al. \cite{jacquier2021non}.}

\section*{Methodology}

\begin{figure}[!ht]
  \centering
    \includegraphics[width=16.5cm,keepaspectratio]{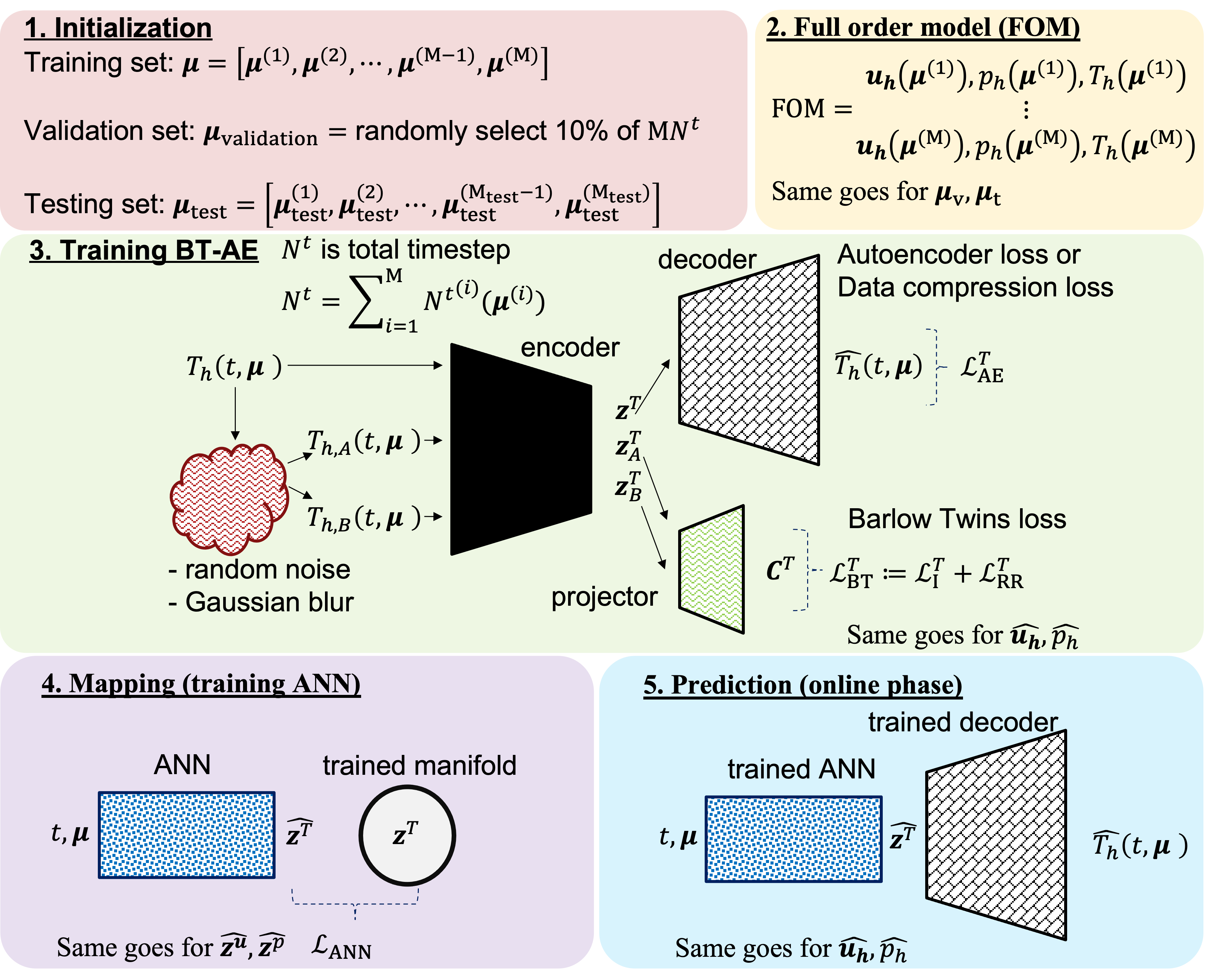}
  \caption{The summary of procedures taken to establish the proposed BT-AE.}
  \label{fig:model}
\end{figure}

A graphical summary of our procedure is presented in Figure \ref{fig:model}: the computations are divided into an offline phase for the ROM construction, which we will show through four consecutive main steps and (single-step) online stage for the ROM evaluation.

The first step of the offline stage represents an initialization of a training set ($\bm{\mu}$), validation set ($\bm{\mu}_{\mathrm{validation}}$), and test set ($\bm{\mu}_{\mathrm{test}}$) of parameters used to train, validate, and test the framework, of cardinality $\mathrm{M}$, $\mathrm{M}_{\mathrm{validation}}$, $\mathrm{M}_{\mathrm{test}}$. For the rest of sections we will discuss only $\bm{\mu}$. The same analogy goes for $\bm{\mu}_{\mathrm{validation}}$ and $\bm{\mu}_{\mathrm{test}}$. Let $\mathbb{P} \subset \mathbb{R}^P$, $P \in \mathbb{N}$, be a compact set representing the range of variation of the parameters $\bm{\mu} \in \mathbb{P}$. For the sake of notation we denote by $\mu_p$, $p = 1, \hdots, P$, the $p$-th component of $\bm{\mu}$. To explore the parametric dependence of the phenomena, we define a discrete training set of $\mathrm{M}$ parameter instances. Each parameter instance in the training set will be indicated with the notation $\bm{\mu}^{(i)}$, for $i = 1, \hdots, \mathrm{M}$. Thus, the $p$-th component of the $i$-th parameter instance in the training set is denoted by $\mu_p^{(i)}$ in the following. The choice of the value of $\mathrm{M}$, as well as the sampling procedure from the range $\mathbb{P}$, is typically user- and problem-dependent. In this work, we use an equispaced distribution for the training set as done in \cite{kadeethum2021non,kadeethum2021nonTH}. \par

In the second step, we query the FOM, based on the finite element solver proposed and made publicly available in Kadeethum et al. \cite{kadeethum2021locally,kadeethum2021nonTH}, for each parameter $\bm{\mu}$ in the training set. In short, we are interested in gravity driven flow in porous media, and here we briefly describe all the equations used in this study: (1) mass balance and (2) heat advection-diffusion equations. Let $\Omega \subset \mathbb{R}^d$ ($d \in \{1,2,3\}$) denote the computational domain and $\partial \Omega$ denote the boundary. $\bm{X}^{*}$ are spatial coordinates in $\Omega$ (e.g., $\bm{X}^{*}=[x^{*}, y^{*}]$ when $d=2$, which we will focus on throughout this study). The time domain is denoted by $\mathbb{T} = \left(0,\tau\right]$ with $\tau>0$ (i.e., $\tau$ is the final time). Primary variables used in this paper are $\bm{u}^* (\cdot , t^*) : \Omega \times  \mathbb{T} \to \mathbb{R}^d$, which is a vector-valued Darcy velocity (\si{m/s}), $p^* (\cdot ,  t^*) : \Omega \times  \mathbb{T} \to \mathbb{R}^d$, which is a scalar-valued fluid pressure (\si{Pa}), and $T^* (\cdot , t^*) : \Omega \times \mathbb{T} \to \mathbb{R}^d$, which is a scalar-valued fluid temperature (\si{C}). Time is denoted as $t^*$ (\si{s}). 

Following Joseph \cite{joseph2013stability}, the Boussinesq approximation to the mass balance equations results in the density difference only appearing in the buoyancy term. The mass balance equation reads

\begin{equation} \label{eq:mass_dim}
\bm{u}^{*}+\bm{\kappa}\left(\nabla p^{*}+{\mathbf{y}}\left(\rho-\rho_{0}\right) g\right)=0,
\end{equation}

\noindent
and

\begin{equation} \label{eq:div_0_dim}
\nabla \cdot \bm{u}^{*}=0
\end{equation}

\noindent
where $\bm{\kappa}=\bm{k}/\mu_f$ is the porous medium conductivity, $\bm{k}$ is the matrix permeability tensor, $\mu_f$ is the fluid viscosity, $\mathbf{y}$ is a unit vector in the direction of gravitational force, $g$ is the constant acceleration due to gravity, $\rho$ and $\rho_0$ are the fluid density at current and initial states, respectively. We assume that $\rho$ is a linear function of $T^{*}$ \cite{chen2006computational, zhang2016mixed}

\begin{equation} \label{eq:rho}
\rho=\rho_{0}\left(1-\alpha\left(T^{*}-T_{0}^{*}\right)\right),
\end{equation}

\noindent
where $\alpha$ is the thermal expansion coefficient, and $T_{0}^{*}$ is the reference fluid temperature. We note that Equation \eqref{eq:rho} is the simplest approximation, and one may easily adapt the proposed method when employing a more complex relationship provided in \cite{lake2014fundamentals}. The heat advection-diffusion equation defined as

\begin{equation} \label{eq:temp_dim}
\gamma \frac{\partial T^{*}}{\partial t^{*}}+\bm{u}^{*} \cdot \nabla T^{*}-K \nabla^{2} T^{*} - {f_c}^* = 0.
\end{equation}

\noindent
Here, $\gamma$ is the ratio between the porous medium heat capacity and the fluid heat capacity, $K$ is the effective thermal conductivity, and ${f_c}^*$ is a sink/source. We follow Nield and Bejan \cite{nield2006convection} and define dimensionless variables as follows

\begin{equation} \label{eq:dim_less}
\bm{X}:=\frac{1}{H} \bm{X}^{*}, \quad t:=\frac{\kappa}{\mu \gamma H^{2}} t^{*}, \quad p:=\frac{\kappa}{K} p^{*}, \quad \bm{u}:=\frac{H}{K} \bm{u}^{*}, \quad T:=\frac{T^{*}-T_{0}^{*}}{\Delta T^{*}}, \quad f_c:=\frac{t^*}{\Delta T^{*}}{f_c}^*,
\end{equation}

\noindent
where $H$ is the dimensional layer depth, and $\Delta T^{*}$ is the temperature difference between two boundary layers. From these dimensionless variables, we could rewrite our Equations \eqref{eq:mass_dim} and \eqref{eq:div_0_dim} as

\begin{equation} \label{eq:mass_dimless}
\begin{split}
\bm{u}+\nabla p-\mathbf{y} \operatorname{Ra} T=0, &\text { \: in \: } \Omega \times \mathbb{T}, \\ 
\nabla \cdot \bm{u}=0,  &\text { \: in \: } \Omega \times \mathbb{T}, \\ 
p=p_{D} &\text { \: on \: } \partial \Omega_{p} \times \mathbb{T}, \\ 
\bm{u} \cdot \mathbf{n}=q_{D} &\text { \: on \:} \partial \Omega_{q} \times \mathbb{T}, \\ 
p=p_{0} &\text { \: in \: } \Omega \text { at } t = 0,
\end{split}
\end{equation}

\noindent
where $\partial \Omega_{p}$ and $\partial \Omega_{q}$ are the pressure and flux boundaries (i.e., Dirichlet and Neumann boundary conditions), respectively. Here, $\mathrm{Ra}$ is the Rayleigh number

\begin{equation} \label{eq:Ra}
\mathrm{Ra}:=\frac{g \alpha \kappa \Delta T^{*} H}{K}.
\end{equation}

\noindent
We then write Equation \eqref{eq:temp_dim} in dimensionless form as follows

\begin{equation} \label{eq:temp_dimless}
\begin{split}
\frac{\partial T}{\partial t}+\bm{u} \cdot \nabla T-\nabla^{2} T - f_c=0,  &\mbox{ \: in \: } \Omega \times (0,\mathbb{T}], \\  
T = T_D  &\mbox{ \: on \: } \partial \Omega_{{T}} \times (0,\mathbb{T}], \\ 
(-\bm{u} T+\nabla T) \cdot \mathbf{n}={T_{\rm in}} \bm{u}\cdot \mathbf{n}  &\mbox{ \: on \: } \partial \Omega_{{\rm in}} \times (0,\mathbb{T}],  \\ 
\nabla T \cdot \mathbf{n} = 0  &\mbox{ \: on \: } \partial \Omega_{{\rm out}} \times (0,\mathbb{T}], \\ 
T={T_{0}}  &\text{ \: in \: } \Omega \text { at } t = 0,
\end{split}
\end{equation}

\noindent
where $\partial \Omega_{{T}}$ is temperature boundary (Dirichlet boundary condition), $\partial \Omega_{\rm in}$ and $\partial \Omega_{\rm out}$ denote inflow and outflow boundaries, respectively, which are defined as

\begin{equation}
\partial \Omega_{\rm in} := \{ \bm{X} \in \partial \Omega : \bm{u} \cdot \mathbf{n} < 0\} \quad \mbox{ and } \quad \partial \Omega_{\rm out} := \{ \bm{X} \in \partial \Omega : \bm{u} \cdot \mathbf{n} \geq 0\}.
\label{eq:in_and_out}
\end{equation}

The detail of discretization could be found in Kadeethum et al. \cite{kadeethum2021locally,kadeethum2021nonTH}, and the FOM source codes are provided in Kadeethum et al. \cite{kadeethum2021locally}. After the second step, we have $\mathrm{M}$ snapshots of FOM results associated with the different parametric configurations in $\bm{\mu}$. Since the problem formulation is time-dependent, the output of the FOM solver for each parameter instance $\bm{\mu}^{(i)}$ collects the time series representing the time evolution of the primary variables for each time-step $t$. Thus, each snapshot contains approximations of the primary variables ($\bm{u}_{h}$, $p_h$, and $T_h$) at each time-step of the partition of the time domain $\mathbb{T}$. Therefore, based on the training set cardinality $\mathrm{M}$ and the number $N^t$ of time-steps, we have a total of $N^t \mathrm{M}$ training data to be employed in the subsequent steps. We note that as our finite element solver utilizes an adaptive time-stepping \cite{kadeethum2021locally,kadeethum2021nonTH}, each snapshot may have a different number of time-steps $N^t$, i.e. $N^t = N^t(\bm{\mu})$. \par 

The third step aims to compress the information provided by the training snapshots provided by the second step. Kadeethum et al. \cite{kadeethum2021nonTH} provide detailed derivations and comparisons between linear and nonlinear compression. Especially the convolutional layers, in their classical form, could not deal with an unstructured data structure (unstructured mesh), which is very common in scientific computing or, more specifically, finite element analysis. Hence, our goal is to develop a nonlinear compression that (1) consistently outperforms (or at least delivers similar accuracy) the linear compression and (2) is compatible with an unstructured data structure. \par

To achieve this goal, we propose a nonlinear compression utilizing feedforward layers in combination with self-supervised learning (SSL) of Barlow Twins (BT) ( Figure \ref{fig:model}). The BT for redundancy reduction is proposed by Zbontar et al. \cite{zbontar2021barlow}. It operates on a joint embedding of noisy images by producing two distorted images from an original one through a series of random cropping, resizing, horizontal flipping, color jittering, converting to grayscale, Gaussian blurring, and solarization. Since we do not operate on structured data (image) but unstructured data produced by finite element solver, we only employ random noise and Gaussian blur operations to produce our noisy data set, see Figure \ref{fig:model}. \par

Let ${z}_1^{\bm{u}}, \cdots, {z}_\mathrm{Q}^{\bm{u}}$, ${z}_1^p, \cdots, {z}_\mathrm{Q}^p$, and ${z}_1^T, \cdots, {z}_\mathrm{Q}^T$ be the nonlinear manifolds of the $\bm{u}_{h}$, $p_h$, and $T_h$, respectively. For the sake of compactness, we will only discuss primary variable $T_h$. The same procedure holds for $\bm{u}_{h}$ and $p_h$. Our goal is to achieve $\mathrm{Q} \ll \mathrm{M} N^t$ where $\mathrm{M} N^t$ is the total training data, which implies that our nonlinear manifolds could represent our training data using much lower dimension. We employ a vanilla AE (using only feedforward layers) that is regularized by Barlow Twins SSL to obtain $\bm{z}^T = \left[ {z}_1^T, \cdots, {z}_\mathrm{Q}^T \right]$. We do not use any batch normalization or dropout. The summary of the training process is presented in Algorithm \ref{al:ae-bt}. We will provide the detailed implementation in \url{https://github.com/sandialabs}. \par

\begin{algorithm}[!ht]
\caption{Training autoencoder (AE) with Barlow Twins (BT) regularization}\label{al:ae-bt}
\# Integrate BT into AE architecture to regularize AE's latent spaces $\bm{z}^T$ \newline
\#\#\# \newline
\# Training data $T_h$ and distorted data $T_{h, A}$, $T_{h, B}$ are input of {encoder}  \newline
\# latent spaces $\bm{z}^T$, $\bm{z}_A^T$, and $\bm{z}_B^T$ are output {encoder}  \newline
\#\#\# \newline
\# latent space $\bm{z}^T$ is output of {decoder}  \newline
\# Approximation of $T_h$, i.e., $\widehat{T_h}$ is output of {decoder}  \newline
\#\#\# \newline
\# latent spaces $\bm{z}_A^T$ and $\bm{z}_B^T$ are input of
{projector} \newline
\# cross-correlation matrix $\mathbf{C}^{T}$ is output of {projector} \newline
\#\#\# 
\begin{algorithmic}[1]
\State Initialize (or load pre-trained models) {encoder}, {decoder}, and {projector} \Comment{size of latent space $\mathrm{Q}$ has to be specified.}
\State Initialize (or load pre-trained optimizers) two optimizers for AE and BT.
\State Load training set $\bm{\mu}$ \Comment{the total training data is $\mathrm{M} N^t$}
\State Randomly select 10\% of $\mathrm{M} N^t$ for validation set $\bm{\mu}_{\mathrm{validation}}$ \Comment{the total training data becomes 90\% of $\mathrm{M} N^t$}
\State Add random noise \Comment{see Equation \eqref{eq:add_noise}}
\State Add Gaussian blur \Comment{see Equation \eqref{eq:gb}}
\State From step 5 and 6, we obtain $T_{h,A}\left(t, \bm{\mu}\right)$ and $T_{h,B}\left(t, \bm{\mu}\right)$ from $T_h\left(t, \bm{\mu}\right)$
\ForEach {epoch}
\State \emph{Outer loop: training BT} \Comment{Batch size $\mathbf{B}_{\mathrm{outer}} = 512$}
\ForEach {$\mathrm{B}_{\mathrm{outer},i}$ $\in$ $\mathbf{B}_{\mathrm{outer}}$}
\State $\bm{z}^T_{h,A}\left(t, \bm{\mu}\right) = \mathrm{encoder}\left(  {T}_{h,A}\left(t, \bm{\mu}\right)  \right)$
\State $\bm{z}^T_{h,B}\left(t, \bm{\mu}\right) = \mathrm{encoder}\left(  {T}_{h,B}\left(t, \bm{\mu}\right)  \right)$
\State $\mathbf{C}^{T}\left(t, \bm{\mu}\right) = \mathrm{projector} \left( \bm{z}^T_{h,A}\left(t, \bm{\mu}\right), \bm{z}^T_{h,B}\left(t, \bm{\mu}\right)  \right)$ 
\State Calculate BT loss $\mathcal{L}_{\mathrm{BT}}^{T}$ \Comment{see Equation \eqref{eq:loss_bt}}
\State Back-propagation of BT loss w.r.t. each $\mathrm{encoder}\left( {\mathbf{W}}, {\mathbf{b}}\right)$ and $\mathrm{projector}\left( {\mathbf{W}}, {\mathbf{b}}\right)$
\State Update $\mathrm{encoder}\left( {\mathbf{W}}, {\mathbf{b}}\right)$ and $\mathrm{projector}\left( {\mathbf{W}}, {\mathbf{b}}\right)$ using BT optimizer
\State Update learning rate $\eta_{c}$ of BT optimizer \Comment{see Equation \eqref{eq:learning_rate}}
\State \emph{Inner loop: training AE} \Comment{Batch size $\mathbf{B}_{\mathrm{inner}} = 32$}
\ForEach {$\mathrm{B}_{\mathrm{inner},i}$ $\in$ $\mathbf{B}_{\mathrm{inner}}$}
\State $\bm{z}_h^T\left(t, \bm{\mu}\right) = \mathrm{encoder}\left(  {T}_h\left(t, \bm{\mu}\right)  \right)$
\State $\widehat{{T}_h}\left(t, \bm{\mu}\right) = \mathrm{decoder}\left(  \bm{z}_h^T\left(t, \bm{\mu}\right)  \right)$
\State Calculate AE loss ${\mathcal{L}_\mathrm{AE}^{T}}$ (data compression loss)  \Comment{see Equation \eqref{eq:loss_ae}}
\State Back-propagation of AE loss w.r.t. each $\mathrm{encoder}\left( {\mathbf{W}}, {\mathbf{b}}\right)$ and $\mathrm{decoder}\left( {\mathbf{W}}, {\mathbf{b}}\right)$
\State Update $\mathrm{encoder}\left( {\mathbf{W}}, {\mathbf{b}}\right)$ and $\mathrm{decoder}\left( {\mathbf{W}}, {\mathbf{b}}\right)$ using AE optimizer
\State Update learning rate $\eta_{c}$ of AE optimizer \Comment{see Equation \eqref{eq:learning_rate}}
\EndFor
\EndFor
\EndFor
\end{algorithmic}
It is noted that, we here only discuss primary variable $T_h$ for the sake of compactness. The same procedures are hold for $\bm{u}_{h}$ and $p_h$. Moreover, this algorithm only reflects the third step in Figure \ref{fig:model}.
\end{algorithm}

\noindent
In short, during the training phase, our BT-AE model is composed of one encoder, one decoder, and one projector. The training entails two sub-tasks; the first is BT (encoder and projector), which takes place in the outer loop. The second sub-task is responsible for the training of AE (encoder and decoder), which takes place inside the inner loop. The main reasons for this procedure are two folds. The first reason is Zbontar et al. \cite{zbontar2021barlow} states that the BT works better with large batch sizes. The AE, however, generally requires a small batch size \cite{wang2017closer,karras2017progressive}. Our previous numerical experiments based on DC-AE \cite{kadeethum2021nonTH} also align with this statement. Consequently, we set our batch size of the outer loop as $\mathbf{B}_{\mathrm{outer}} = 512$, and the batch size of the inner loop as $\mathbf{B}_{\mathrm{inner}} = 32$ \par

Prior to the training, we distort our training set (i.e., creating $T_{h,A}\left(t, \bm{\mu}\right)$ and $T_{h,B}\left(t, \bm{\mu}\right)$ from $T\left(t, \bm{\mu}\right)$) through a series of two operations. First, \emph{add random noise} is added as follows

\begin{equation}\label{eq:add_noise}
\widetilde{T_{h,A}}\left(t, \bm{\mu}\right), \widetilde{T_{h,B}}\left(t, \bm{\mu}\right) = T\left(t, \bm{\mu}\right) + \epsilon \operatorname{SD}\left(T\left(t, \bm{\mu}\right)\right) \mathcal{G}\left(0,1\right)
\end{equation}

\noindent
where $\widetilde{T_{h,A}}\left(t, \bm{\mu}\right), \widetilde{T_{h,B}}\left(t, \bm{\mu}\right)$ are intermediate distorted input data. The constant $\epsilon$, which is set to $0.1$, determines the noise level as it is multiplied with the standard deviation of the input field. $\mathcal{G}\left(0,1\right)$ is a random value which is sampled from the standard normal distribution with mean and standard deviation of zero and one, respectively. 

Subsequently, we pass $\widetilde{T_{h,A}}\left(t, \bm{\mu}\right), \widetilde{T_{h,B}}\left(t, \bm{\mu}\right)$ through Gaussian blur operation, which reads

\begin{equation}\label{eq:gb}
{T_{h,A}}\left(t, \bm{\mu}\right), {T_{h,B}}\left(t, \bm{\mu}\right)=\frac{1}{\sqrt{2 \pi \operatorname{SD}\left(\widetilde{T_{h,A}}\left(t, \bm{\mu}\right), \widetilde{T_{h,B}}\left(t, \bm{\mu}\right)\right)^{2}}} \mathrm{exp}\left({-\frac{{\widetilde{T_{h,A}}\left(t, \bm{\mu}\right), \widetilde{T_{h,B}}\left(t, \bm{\mu}\right)}^{2}}{2 \operatorname{SD}\left(\widetilde{T_{h,A}}\left(t, \bm{\mu}\right), \widetilde{T_{h,B}}\left(t, \bm{\mu}\right)\right)^{2}}}\right)
\end{equation}

\noindent
to obtain $T_{h,A}\left(t, \bm{\mu}\right)$ and $T_{h,B}\left(t, \bm{\mu}\right)$. 

We use a number of the epoch of 50, see Algorithm \ref{al:ae-bt}. The outer loop works as follows: training BT begins with passing $T_{h,A}\left(t, \bm{\mu}\right)$ and $T_{h,B}\left(t, \bm{\mu}\right)$ to the encoder (it is noted we have only one encoder) resulting in $\bm{z}_A^T\left(t, \bm{\mu}\right)$ and $\bm{z}_B^T\left(t, \bm{\mu}\right)$. We then use $\bm{z}_A^T\left(t, \bm{\mu}\right)$ and $\bm{z}_B^T\left(t, \bm{\mu}\right)$ as input to the projector resulting in cross-correlation matrix $\mathbf{C}^{T}\left(t, \bm{\mu}\right)$. $\mathbf{C}^{T}\left(t, \bm{\mu}\right)$ is a square matrix with the dimensionality of the projector's output, and its values range between -1, perfect anti-correlation, and 1, perfect correlation.

The Barlow Twins loss $\mathcal{L}_{\mathrm{BT}}^{T}$ (BT loss) is then calculated using

\begin{equation} \label{eq:loss_bt}
\mathcal{L}_{\mathrm{BT}}^{T} := \mathcal{L}_{\mathrm{I}}^{T} + \mathcal{L}_{\mathrm{RR}}^{T}
\end{equation}

\noindent
where
\begin{equation} \label{eq:loss_I}
\mathcal{L}_{\mathrm{I}}^{T} := \sum_{i}\left(1-\mathbf{C}_{i i}^T\left(t, \bm{\mu}\right)\right)^{2},
\end{equation}

\noindent
and
\begin{equation} \label{eq:loss_RR}
\mathcal{L}_{\mathrm{RR}}^{T} := \lambda \sum_{i} \sum_{j \neq i} {\mathbf{C}_{i j}^T\left(t, \bm{\mu}\right)}^{2},
\end{equation}

\noindent
where $\mathbf{C}_{i i}^T\left(t, \bm{\mu}\right)$ denotes the $i$-th diagonal entry of $\mathbf{C}^{T}\left(t, \bm{\mu}\right)$, $\lambda$ is a positive constant, which is set to $5 \times 10^{-3}$ as recommended by Zbontar et al. (2021) \cite{zbontar2021barlow}, and $\mathbf{C}_{i j}^T$ are off-diagonal entries of $\mathbf{C}^{T}$. In short, we train our BT part by trying to force $\mathcal{L}_{\mathrm{I}}^{T}$ to 1, but $\mathcal{L}_{\mathrm{RR}}^{T}$ to 0 resulting in teaching the encoder and projector learn how to get rid off noise from the distorted data, $T_{h,A}\left(t, \bm{\mu}\right)$ and $T_{h,B}\left(t, \bm{\mu}\right)$, and construct a representation that conserves as much $T\left(t, \bm{\mu}\right)$ information as possible. \par

Here, we follow the training procedures used by Kadeethum et al. \cite{kadeethum2021framework, kadeethum2021nonTH}. We use the ADAM algorithm\cite{kingma2014adam} to adjust learnable parameters of encoder($\mathrm{W}$ and $\mathrm{b}$) and projector($\mathrm{W}$ and $\mathrm{b}$). The learning rate ($\eta$) is calculated as \cite{loshchilov2016sgdr}

\begin{equation}\label{eq:learning_rate}
\eta_{c}=\eta_{\min }+\frac{1}{2}\left(\eta_{\max }-\eta_{\min }\right)\left(1+\cos \left(\frac{\mathrm{step_c}}{\mathrm{step_f}} \pi\right)\right)
\end{equation}

\noindent
where $\eta_{c}$ is a learning rate at step $\mathrm{step_c}$, $\eta_{\min }$ is the minimum learning rate, which is set as $1 \times 10^{-16}$, $\eta_{\max }$ is the maximum or initial learning rate, which is selected as $1 \times 10^{-4}$, $\mathrm{step_c}$ is the current step, and $\mathrm{step_f}$ is the final step. We note that each step refers to each time we perform back-propagation, including updating both encoder and projector's parameters. \par


The inner loop is as follows: training AE starts with obtaining $\bm{z}^T\left(t, \bm{\mu}\right)$ by passing $T_h\left(t, \bm{\mu}\right)$ to the encoder. We then use $\bm{z}^T\left(t, \bm{\mu}\right)$ to reconstruct $\widehat{T}_h\left(t, \bm{\mu}\right)$ through the decoder. Subsequently, we calculate our data compression loss or AE loss ($\mathcal{L}_\mathrm{AE}^{T}$) using 

\begin{equation} \label{eq:loss_ae}
{\mathcal{L}_\mathrm{AE}^{T}} := 
    {\mathrm{MSE}^{T}} = 
    \frac{1}{\mathrm{M} N^t} \sum_{i=1}^{\mathrm{M}}\sum_{k=0}^{N^t}\left|\widehat{T}_h\left(t^k, \bm{\mu}^{(i)}\right)-T_h\left(t^k, \bm{\mu}^{(i)}\right)\right|^{2}.
\end{equation}

\noindent
Similar to the training of BT, we use ADAM to adjust learnable parameters of encoder($\mathrm{W}$ and $\mathrm{b}$) and decoder($\mathrm{W}$ and $\mathrm{b}$) according the gradient of Equation \eqref{eq:loss_ae}. The $\eta_{c}$ is adjusted by Equation \eqref{eq:learning_rate}. In contrast to the training of BT, we use $\eta_{\min } = 1 \times 10^{-16}$, and $\eta_{\max } = 1 \times 10^{-5}$.

Following the training of the BT-AE, we now establish the manifold $\bm{z}^T\left(t, \bm{\mu}\right), \quad \forall t \in \mathbb{T} \: \text{and} \: \forall\bm{\mu} \in \mathbb{P}$ during the fourth step shown in Figure \ref{fig:model}. The data available for this task are the pairs $(t, \bm{\mu})$ and $\bm{z}^T\left(t, \bm{\mu}\right)$ in the training set. We achieve this through the training of artificial neural networks (ANN). Following Kadeethum et al. \cite{kadeethum2021non, kadeethum2021nonTH}, our ANN has five hidden layers, and each layer has seven neurons. We use tanh as our activation function. Here, we use a mean squared error (${\mathrm{MSE}^{\bm{z}^T}}$) as the metric of our network loss function, defined as follows

\begin{equation}\label{eq:loss_ann_nonlinear}
{\mathcal{L}_\mathrm{ANN}^{T}} := {\mathrm{MSE}^{\bm{z}^T}}=\frac{1}{\mathrm{M} N^t} \sum_{i=1}^{\mathrm{M}}\sum_{k=0}^{N^t}\left|\widehat{\bm{z}}^T\left(t^k, \bm{\mu}^{(i)}\right)-{\bm{z}^T}\left(t^k, \bm{\mu}^{(i)}\right)\right|^{2}.
\end{equation}

\noindent
To minimize Equation \eqref{eq:loss_ann_nonlinear}, we use the ADAM algorithm to adjust each neuron $\mathrm{W}$ and $\mathrm{b}$, a batch size of 32, a learning rate of 0.001, a number of epoch of 10,000, and we normalize both our input ($t, \bm{\mu}$) and output ($\bm{z}^T$) to $[0, 1]$. To prevent our networks from overfitting behavior, we follow early stopping and generalized cross-validation criteria \cite{hesthaven2018non,prechelt1998early,prechelt1998automatic}. Note that instead of literally stopping our training cycle, we only save the set of trained weight and bias to be used in the online phase when the current validation loss is lower than the lowest validation from all the previous training cycle. \par

During the online phase (the fifth step shown in Figure \ref{fig:model}), we utilize the trained ANN and the trained decoder to approximate $\widehat{{T}}_{h}\left(\cdot; t, \bm{\mu}\right)$ for each inquiry (i.e., a pair of $(t, \bm{\mu})$ ) through 

\begin{equation}
\widehat{\bm{z}}^T\left(\cdot; t, \bm{\mu}\right) = \operatorname{ANN} \left( t, \bm{\mu} \right),
\label{eq:online_solution_z_nonlinear}
\end{equation}

\noindent
and, subsequently, 

\begin{equation}
\widehat{{T}}_{h}\left(\cdot; t, \bm{\mu}\right) = \operatorname{decoder} \left(\widehat{\bm{z}}^{T}\left(\cdot; t, \bm{\mu}\right) \right).
\label{eq:online_solution_nonlinear}
\end{equation}

\rev{We note that, for the prediction phase, our ROM could be evaluated using any timestamps, including one that does not exist in the training phase (i.e., any $t$ that lies within $[t^{0}, \tau$]) because our ROM treats the time domain $\mathbb{T}$ continuously. Besides, in contrast with the FOM, the ROM is not bound by the CFL condition and can predict the quantities of interest at any specific time without intermediate computation. Hence, our proposed framework can reduce the computational time significantly.}

\bibliography{cites}

\section*{Acknowledgements}
TK and HY were supported by the Laboratory Directed Research and Development program at Sandia National Laboratories and US Department of Energy Office of Fossil Energy and Carbon Management, Science-Informed Machine Learning to Accelerate Real Time Decisions-Carbon Storage (SMART-CS) initiative. 
FB thanks the project ``Numerical modeling of flows in porous media'' funded by the Catholic University of the Sacred Heart, and the European Union's Horizon 2020 research and innovation program under the Marie Skłodowska-Curie Actions, grant agreement 872442 (ARIA).
DO acknowledges support from Los Alamos National Laboratory's Laboratory Directed Research and Development Early Career Award (20200575ECR). 
YC acknowledges LDRD funds (21-FS-042) from Lawrence Livermore National Laboratory (LLNL-JRNL-831095). Lawrence Livermore National Laboratory is operated by Lawrence Livermore National Security, LLC, for the U.S. Department of Energy, National Nuclear Security Administration under Contract DE-AC52-07NA27344.
NB acknowledges startup  support from the Sibley School of Mechanical and Aerospace Engineering, Cornell University.
Sandia National Laboratories is a multimission laboratory managed and operated by National Technology and Engineering Solutions of Sandia LLC, a wholly owned subsidiary of Honeywell International Inc. for the U.S. Department of Energy’s National Nuclear Security Administration under contract DE-NA0003525. This paper describes objective technical results and analysis. Any subjective views or opinions that might be expressed in the paper do not necessarily represent the views of the U.S. Department of Energy or the United States Government.

\section*{Author contributions statement}

\textbf{T. Kadeethum}: Conceptualization, Formal analysis, Software, Validation, Writing - original draft, Writing - review \& editing. \textbf{F. Ballarin}: Conceptualization, Formal analysis, Supervision, Validation, Writing - review \& editing. \textbf{D. O'Malley}: Conceptualization, Formal analysis, Supervision, Validation, Writing - review \& editing. \textbf{Y. Choi}: Conceptualization, Formal analysis, Supervision, Validation, Writing - review \& editing. \textbf{N. Bouklas}: Conceptualization, Formal analysis, Funding acquisition, Supervision, Writing - review \& editing. \textbf{H. Yoon}: Conceptualization, Formal analysis, Funding acquisition, Supervision, Writing - review \& editing.

\section*{Competing interests}
The authors declare no competing interests

\section*{Code and data availability}

Our model scripts and all data generated or analyzed during this study will be available publicly through the Sandia National Laboratories software portal — a hub for GitHub-hosted open source projects (\url{https://github.com/sandialabs}). 


\end{document}